\documentclass[prl,reprint,showpacs]{revtex4-1}
\usepackage{times,amsmath,graphicx,latexsym}

\begin{document}

\title{Exact Solution for Vortex Dynamics in Temperature Quenches of Two-Dimensional Superfluids}

\author{Andrew Forrester, Han-Ching Chu,  and Gary A. Williams}
\affiliation{Department of Physics and Astronomy, University of
California,
Los Angeles, CA 90095}
\date{\today}
\begin{abstract}
An exact analytic solution for the dynamics of vortex pairs is obtained for rapid temperature quenches of a superfluid film starting from the line of critical points below the critical temperature $T_{KT}$.  An approximate solution for quenches at and above above  $T_{KT}$ provides insights into the origin of logarithmic transients in the vortex decay, and is in general agreement with recent simulations of the quenched XY model.   These results confirm that there is no ``creation" of vortices whose density increases with the quench rate as predicted by the Kibble-Zurek theory, but only monotonic decay of the thermal vortices already present at the initial temperature.
\end{abstract}


\maketitle

Although the phase-ordering kinetics of temperature-quenched thermodynamic systems have been studied for many decades \cite{bray}, there are only a few exact results for the dynamics of the recovery to equilibrium \cite{oned,brayrutenberg}. For many systems, progress in the field has been made by asserting that dynamic scaling should apply to a quenched system, that for a system with non-conserved order parameter the dynamics will be characterized by a growing length scale $\xi (t) = \xi_0 t^{1/z}$, where $z \approx 2$ is the dynamical exponent of model A in the classification of Halperin and Hohenberg \cite{halperin} and $t$ is the time from the quench to low temperature.  Scaling holds if solutions involving a length scale $r$ only depend on the ratio $r/\xi$.  The growing length scale characterizes the domain growth of the topological defects of the order parameter as the system becomes completely ordered at long times.  A phenomenological argument \cite{bray,toyoki} is commonly used to predict the time dependence of the decaying defect density:
\begin{equation}
\rho(t)  \propto \xi ^{ - n}  \propto t^{ - n/z}
\label{eq1}  
\end{equation}
where $n$ is the number of components of the order parameter.  Superfluids are in the $n = 2$ universality class, where the defects are quantized vortices, so if $z  = 2$ then dynamic scaling predicts a $1/t$ decay of the vortex density.  Computer simulations of spin systems with varying $n$ gave general agreement with Eq.\,(\ref{eq1}), though only at long time and length scales \cite{bray2}.  

However, a problematic case for dynamic scaling has been two-dimensional superfluids, where the defects are the vortex pairs of the Kosterlitz-Thouless theory \cite{kt} characterizing the equilibrium phase transition occurring at the critical temperature $T_{KT}$.  The above arguments would give a vortex density decaying as $t^{-1}$ for quenches from well above $T_{KT}$ to very low temperatures.  Simulations of the XY model \cite{mondello2d,yurke,cugliandolo}, however, showed the initial vortex decay to be considerably slower than this, and then only at very long times finally approached the predicted exponent of $-1$.  The behavior could be modeled as a
$\ln t/t$ variation, but this requires altering the dynamic length scale to vary as $(t/\ln t)^{1/z}$ for initial temperatures above $T_{KT}$.  This change has been cited \cite{braybriant} as a breakdown of dynamic scaling, though others \cite{rutenberg} find such a sudden change in the dynamic scale still fully consistent with scaling.  To further complicate the issue, numerical solutions of the Fokker-Planck equation for vortex pairs carried out by two of us \cite{hcchu} for quenches starting from $T_{KT}$ showed a rather different logarithmic form for the decay, as $1/(t\ln t)$, and quenches from initial temperatures below $T_{KT}$ showed a temperature-dependent decay that became considerably more rapid than an exponent of $-1$ as the starting temperature was reduced. 

Here we show that an exact solution to the Fokker-Planck equation for instantaneous quenches can explain much of the apparently anomalous behavior listed above.  The solution yields a key insight that the quench dynamics depends crucially on the vortex-pair distribution function at the initial temperature $T_i$.  The logarithmic corrections to the vortex decay at $T_i$ = $T_{KT}$ are found to arise from logarithmic corrections to the power-law behavior of the initial distributions, which in turn comes from the increasing renormalization of the superfluid density near $T_{KT}$.

The starting point is the Fokker-Planck equation for the vortex pair distribution function \cite{ahns} as used in Ref.\,\cite{hcchu},
\begin{equation}
\frac{{\partial \,\Gamma }}{{\partial \,t}} = \frac{1}{r}\,\;\frac{\partial }{{\partial r}}\left( {r\frac{{\partial \Gamma }}{{\,\partial r}} + 2\pi K\,\Gamma} \right)
\label{eq2}
\end{equation}
where $\Gamma (r,t)$ is the distribution function for pairs of separation $r$ at time $t$ from the instantaneous quench, with $r$ measured in units of the vortex core radius $a_0$, $t$ in units of the diffusion time $a_0^{2}/2D$ with $D$ the vortex diffusion constant, and $\Gamma$ in units $a_0^4$.  In the limit of the very low vortex densities that we consider, a term quadratic in 
$\Gamma$  corresponding to the recombination of opposite-sign vortices on differing vortex pairs \cite{ahns} can be neglected in 
Eq.(\ref{eq2}).  Only the recombination of same-pair vortices at the closest separation $r =1$ plays a role in the dynamics in this limit.  The factor $K={\hbar ^2\sigma _s}$/${m^2k_BT}$ is the dimensionless areal 
superfluid density.  The starting value of $\Gamma$ at $r = 1$ and $t = 0$ is $\Gamma_0 = \exp(-\pi^2 K_{0i}/ 2)$, using the Villain approximation for the vortex core energy.  At longer length scales $K$ is renormalized from its initial value $K_{0i}$ by the Kosterlitz scaling relation \cite{kt}
\begin{equation}
\frac{{\partial K}}{{\partial r}} =  - 4\pi ^3 r^3 K^2 \Gamma \quad.
\label{eq3}     
\end{equation}
We note that these equations for the vortex dynamics have been well verified in finite-frequency experiments on helium films \cite{reppy}.  The most recent experiments \cite{hieda} have driven the films far from equilibrium at very high frequencies up to 60 MHz, resulting in strong broadening effects at the transition in agreement with the theory.

 At $t = 0$ the initial condition is that the system is equilibrated to a thermal bath at the temperature $T_i$, corresponding to the initial $K_i$ as determined from Eq.\,(\ref{eq3}).  For quenches starting from below $T_{KT}$ we can make the approximation that $K_i$ is effectively a constant, since the spatial renormalization from Eq.\,(\ref{eq3}) is rapid, changing from the initial $K_{0i}$ to the renormalized $K_i$ over a length scale that can be less than a core radius.  The equilibrium distribution then varies as $\Gamma_0 ' \,r^{-2 \pi K_i}$, where $\Gamma_0 '$ is slightly smaller than $\Gamma_0$ due to the renormalization.  Since the KT transition is a line of critical points, all quenches starting at $T_i  \le T_{KT}$ are critical quenches \cite{brayrutenberg}.

Immediately after the quench, $K$ in Eq.\,(\ref{eq3}) takes on the new much larger value $K_f$ corresponding to the low final temperature $T_f$.   Again, for quenches starting from below $T_{KT}$ the time dependence of $K_f$ is minimal (as well as the space dependence), recovering after the quench to a superfluid fraction of one in less than a diffusion time \cite{hcchu}.  The thermal bath now acts as a delta-function sink to absorb the out-of-equilibrium smallest pairs at $r = 1$ as they annihilate there, which goes on until the pair density falls to the equilibrium value at $T_f$ (which for a quench to e.g. $T_f  = 0.1 \,T_{KT}$ can be 20 orders of magnitude smaller than the starting value).

In the limit of constant $K_i$ and $K_f$, we can solve Eq.\,(\ref{eq2}) for $t  > 0$ by separation of variables, as detailed in the Supplemental Materials.  The solution for the distribution function becomes
\begin{equation}
\Gamma (r,t) = \beta \,\,_1 F_1 \left[ {\pi K_i ,1 + \pi K_f , - \frac{{r^2 }}{{2z{\kern 1pt} \,t^{2/z} }}} \right]\;\,t^{ - 2\pi K_i /z} 
\label{eq4}
\end{equation}
where $_1 F_1$ is the confluent hypergeometric function of the first kind, $z$ = 2 exactly, and
\begin{equation}
\beta  = \Gamma _0 ' \left( {\frac{{\:G (1 + \pi K_f  - \pi K_i )}}{{(2z)^{\pi K_i } \;G (1 + \pi K_f )}}} \right)
\label{eq5}
\end{equation}
with $G$ the Euler gamma function.   
This distribution is evaluated in Fig.\,1a for a quench from $T_i = 0.9 \;T_{KT}$ to $T_f = 0.1 \;T_{KT}$, corresponding to parameter values $\Gamma_{0} ' = 2.576\times 10^{-4}$, $K_i  =  0.814$, and $K_f  =  7.479$.  With no adjustable parameters this is seen to give a precise description of the numerical results, and the dependence on $r/t^{1/z}$ exactly satisfies dynamic scaling.

The time dependence of the vortex pair density (in units $a_0^2$) is found by integrating the distribution function,
\begin{equation}
\begin{array}{l}
 \rho (t) = \int_1^\infty  {\Gamma (r,t)\,2\pi r} dr \\ 
 \begin{array}{*{20}c}
   {}  \\
   \begin{array}{l}
  = 2 \pi \, \beta \left( {\frac{{z\,(2\pi K_f )}}{{(2\pi K_i  - 2)}}} \right)\:_1 F_1 \left[ {\pi K_i  - 1,\:\pi K_f ,\: - \frac{{t^{ - 2/z} }}{{2z}}} \right] \\ 
  \\ 
 \quad \quad  \times \:t^{ - (2\pi K_i  - 2)/z} \quad. \label{eq6} \\ 
 \end{array}  \\
\end{array} \\ 
 \end{array}
\end{equation}
 \begin{figure}[b]
\begin{center}
\includegraphics[width=0.5\textwidth]{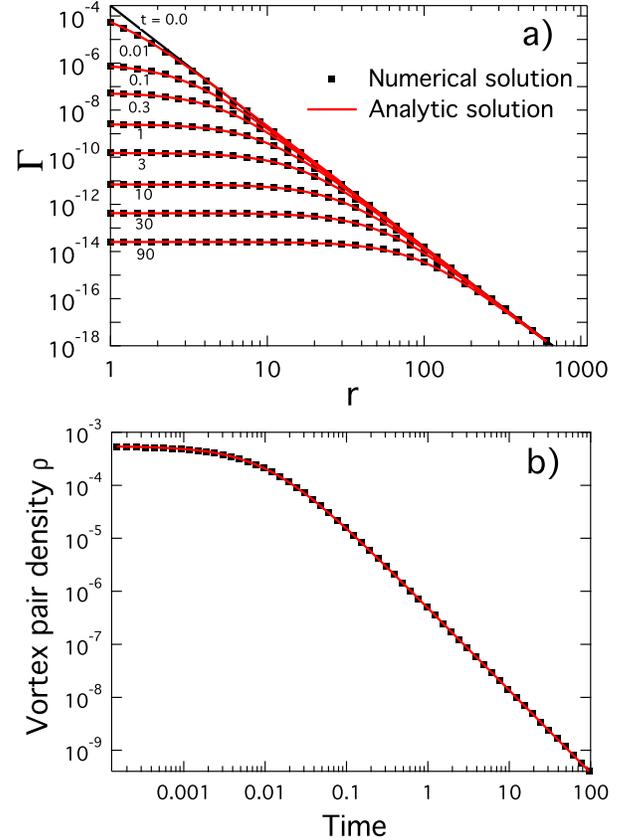} 
\end{center}
\caption{Comparison of the numerical and analytic solutions for the vortex distribution function and vortex pair density as functions of time, for an instantaneous quench from 0.9 $T_{KT}$ to 0.1 $T_{KT}$ . (Color online)}
\label{fig1}
\end{figure}

The hypergeometric function in this case rises rapidly from zero until the time $t \approx 1$ where it becomes constant and equal to one.  Beyond that point the time dependence is then accurately $
t^{ - z_{scale} /z} 
$ where
\begin{equation}
z_{scale}  = 2\pi K_i  - 2 = 4\frac{{\sigma _s (T_i )}}{{\sigma _s (T_{KT} )}}\frac{{T_{KT} }}{{T_i }} - 2
\quad   \label{7}
\end{equation} 
is the dynamic exponent first considered by Minnhagen and co-workers \cite{minn}.   The temperature dependence of $z_{scale}$ is shown in  Fig.\,3 of Ref.\,\cite{hcchu}.  This variation of the vortex density in time is the same as observed earlier, but at that time it was not clear that in fact two dynamic exponents, $z$ and $z_{scale}$, are involved in the solution.  Figure 1b shows the time decay of the density for the same parameters as 1a, where at long times the decay is $t^{-1.557}$. 
\begin{figure}[t]
\begin{center}
\includegraphics[width=0.5\textwidth]{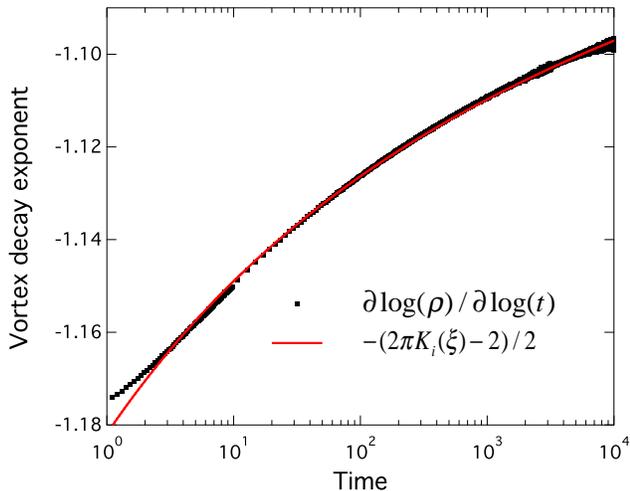}
\end{center}
\caption{Comparison of the numerically computed logarithmic slope of the vortex decay curve at $T_{KT}$ in Ref.\,\cite{hcchu} with the exponent $-z_{scale}/z$ using $K_i(r)$ at the time $t = (r / \xi_0)^2$.  Roundoff error becomes appreciable at long times in the data since $\rho$ becomes very small.   (Color online)}
\label{fig2}\end{figure}

The exact solutions show that Eq.\,(\ref{eq1}) is not correct for quenches starting below $T_{KT}$.  Right at $T_{KT}$, $z_{scale}$ takes the value of 2 at long length scales, and it is certainly possible that this could be related to $n = 2$ of the superfluid universality class.  However, we are unaware of any direct relation between these quantities.
\begin{figure}[t]
\begin{center}
\includegraphics[width=0.5\textwidth]{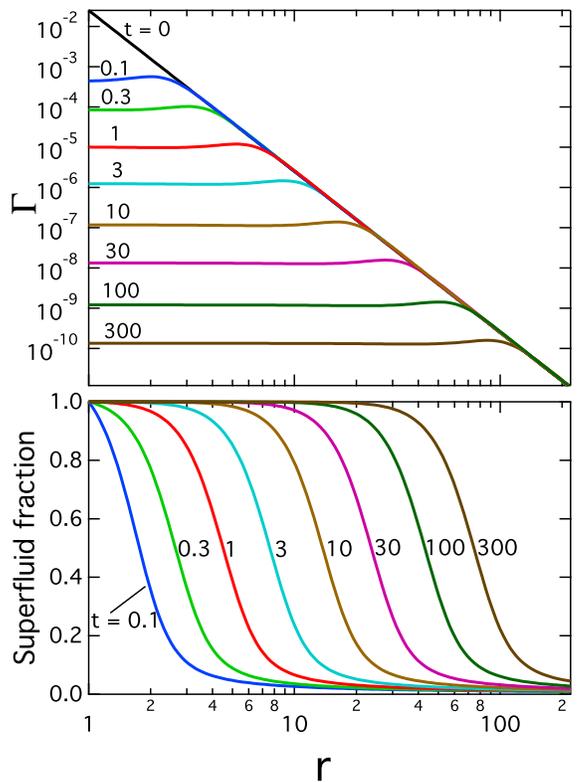} 
\end{center}
\caption{Pair distribution function for a quench from 2.0 to 0.1 $T_{KT}$ (upper plot).  The lower plot shows the time dependence of the scale-dependent superfluid density.   (Color online)}
\label{fig3}
\end{figure}

At initial temperatures higher than about 0.95\,$T_{KT}$ the above analysis begins to fail because the initial vortex distribution is no longer accurately a power law.  Right at $T_{KT}$ there is a strong logarithmic correction to $K_i$ such that $2 \pi K_i$ approaches the KT value of 4 ($K_i = 2/\pi$) only at extremely large pair separations (it is 4.373 at $r = 10$ and 4.197 at $r = 1000$).  It is still possible to proceed if one notes that in Eq.\,(\ref{eq2}) the derivatives are only important near $r  \approx \xi$, since at smaller length scales the distribution is nearly flat, while at larger scales it falls off rapidly.  The solutions of Eqs.\,(\ref{eq4}) and (\ref{eq6}) should remain approximately valid, but now $2 \pi K_i$ has to be regarded as the local value at $r \approx \xi$, i.e.  it becomes  time-dependent.  This in turn means that the vortex density decay exponent $-z_{scale}/z$ will be time-dependent, approaching $-1$ only at very long times.  

We can check this scenario for a quench from $T_{KT}$ by numerically evaluating $K_i$ from Eq.\,(\ref{eq3}) as a function of length scale and computing the decay exponent $-(2 \pi K_i - 2) / 2$ at the corresponding times found from $r = \xi$.  This can then be directly compared to Eq.\,(\ref{eq6}) by  numerically finding the slope of the vortex density decay curve for the quench from $T_{KT}$ in Fig.\,2 of Ref.\,\cite{hcchu} as a function of time, shown as the data points in Fig.\,2 of this paper.  Since $\xi_0$ is not uniquely determined by the above analysis, we have matched the computed exponent (solid curve in Fig.\,2) to the data at one point, arbitrarily taken to be $t = 100$, which yields $\xi_0 = 11.5$.  The agreement over more than three orders of magnitude in $t$ between the data and the computation shows with little doubt that the $1/(t\ln t)$ behavior seen in \cite{hcchu} was entirely due to the log corrections to $K_i$ in the initial distribution.   A scaling plot of $t^{2\pi K_i /z} \,\Gamma$ versus $r / \xi$ for this case (shown in the Supplemental Materials) is found to give accurate data collapse if $\xi$ is taken to be 
$\xi_0 t^{1/z}$, whereas including a log term in $\xi$ does not yield data collapse.

Similar ideas can be applied to quenches from well above $T_{KT}$, though in that case less is known about the vortex distribution, and the initial superfluid density is zero.  We know at $T_ {KT}$ that vortex pairs of extremely large separation ($r \to \infty$)
distributed at random (``free'' vortices) have a pair distribution falling off as $r^{-4}$, and so we expect for temperatures above $T_{KT}$ that this distribution will hold for large but finite $r$.  Fig.\,\ref{fig3} shows a numerical evaluation of  Eqs.\,(\ref{eq2}) and (\ref{eq3}) for an instantaneous quench from 2.0 to 0.1 $T_{KT}$ assuming the initial distribution is $r^{-4}$ at all length scales.  As the smallest pairs decay the scale-dependent superfluid fraction quickly recovers at the smallest scales, and then with increasing time recovers at longer and longer scales as the larger pairs decay.  The distribution function shows a small ``hump'' at the scales where the superfluid fraction is varying most rapidly, which appears very similar to the small ``humps'' seen in the simulations \cite{cugliandolo} of quenches from 2\,$T_{KT}$.  Integration of these distribution curves gives a vortex density decaying accurately as $t^{-1.0}$ for $t >$ 10.

At short length scales above $T_{KT}$ there will still be significant correlations between the vortices.  The pair distribution has been found in the simulations of  Ref.\,\cite{cugliandolo} at 2\,$T_{KT}$ to fall off as $r^{-3.5}$ for $r <$ 50.  Iterating the equations using this form in the quench from  2\,$T_{KT}$ gives results entirely similar to Fig. 3, but now upon integrating the vortex density falls off as $t^{-0.6}$, just about the behavior seen at short times in the simulations.  We are unable to carry out a more complete calculation, however, since we do not know more precisely how the distribution changes over from $r^{-3.5}$ to $r^{-4}$ at large length scales.  A logarithmic change between these two exponents could well give rise to the $\ln t/t$ variation that is observed.  Simulations on much larger lattice sizes will be needed to study the distribution function at large length scales.

The exact solutions presented here for the quench of 2D XY superfluids validate the conclusion in Ref. \cite{hcchu} that there is no ``creation" of vortices in a quench, but only monotonic decay of the existing thermal vortices.  There is no term in the solution that increases the number of vortices, and which might have possibly been missed in the numerical approximations. This is contrary to the predictions of the Kibble-Zurek scenario \cite{kz} that vortices will be created in a superfluid quench to low temperatures, with higher densities appearing the faster the quench rate.  The numerical studies with finite quench times \cite{hcchu} found just the opposite result, that actually more vortices were left over following a slower quench, since in that case the system spends more time at higher temperatures where the thermal vortex density is higher.  The present results verify that no excess vortices are created even for an instantaneous quench from above $T_{KT}$.  The same result has now been seen in XY model simulations \cite{cugliandolo} for quenches starting from 2\,$T_{KT}$, where again there was only monotonic decay of the initial vortex density, and the variation with quench rate was entirely similar to the previous numerical results \cite{hcchu}.  The problem in the Kibble-Zurek argument is the restriction to measuring the vortex density only at the ``freezeout'' sampling time, which increases with the quench time. But since the the pairs continually decay, of course this will always result in lower vortex densities for a longer quench time and hence a later sampling time.  But in fact the vortex densities can be measured at all times $t$, as shown in the results above, and it then becomes quite clear that the instantaneous superfluid quench has the lowest vortex density at all times of any quench rate, since it most rapidly gets to the lowest temperature.

In summary, we propose the first exact solution for the vortex dynamics following a critical quench starting from below $T_{KT}$ in two-dimensional XY superfluids.  The solution highlights the key role of the initial vortex distribution, and this allows a consistent explanation of  the logarithmic deviations at $T_{KT}$ seen in earlier work as being due to logarithmic corrections in the initial distribution.  It will be important to check the validity of the solution by carrying out XY model simulations for quenches starting below $T_{KT}$, looking for the increasingly rapid decay of the vortex density predicted by Eq.\,(\ref{eq6}).  Such simulations could definitively test the link between the Monte Carlo dynamics and the Kosterlitz-Thouless vortex-pair dynamics of Eq.\,(\ref{eq2}).  Experimentally, these predictions for quenches are probably impossible to test in superfluid helium films where the diffusion time is only tens of picoseconds, but may possibly be accessible in the 2D superfluid polariton condensates  \cite{polaritons}, which can be created and the vortices probed at these rapid time scales.

This work is supported by the US National Science Foundation, Grant DMR 09-06467.  We thank L. Cugliandolo, 
A. Jeli\'c, and W. Newman for useful discussions.

\pagebreak
\section*{Supplementary Material}
\subsection*{Exact solution}
In the limit of constant $K_i$ and $K_f$ we can solve Eq.\,(2) of the main text  for $t  > 0$ by separation of variables, writing 
\begin{equation}
\Gamma (r,t) = {F}(w)\,{f}(t)
\end{equation}
where $w = r/t^{1/z}$ is the dynamic scaling variable.  Using the chain rule, the Fokker-Planck equation then becomes
\begin{equation}
\frac{1}{f}\frac{{\partial f}}{{\partial t}}t^{2/z}  = \frac{1}{F}\left[ {\frac{{\partial ^2 F}}{{\partial w^2 }} + \frac{1}{w}\frac{{\partial F}}{{\partial w}}(1 + 2\pi K_f ) + \frac{{w\,t^{2/z} }}{{z\,t}}\frac{{\partial F}}{{\partial w}}} \right]{\rm{ }}\,.
 \label{1}\\ 
 \end{equation}
This equation separates if and only if $z=2$, and taking the separation constant as $-\alpha /z$ yields the two equations
\begin{equation}
\frac{1}{f}\frac{{\partial f}}{{\partial t}} + \frac{{\alpha}}{{z\,t}} = 0
\label{2}
\end{equation}
\begin{equation}
\frac{{\partial ^2 F}}{{\partial w^2 }} + \frac{1}{w}\frac{{\partial F}}{{\partial w}}(1 + 2\pi K_f) + \frac{w}{{z\,}}\frac{{\partial F}}{{\partial w}} + \frac{{\alpha}}{z}F = 0
\label{3}
\end{equation}
These are straightforward to solve, and the solution for (\ref{3}) finite at $w$ = 0 (i.e. $t \to \infty$) is
\begin{equation}
F(w) = \beta \,\,_1 F_1 \left[ {\frac{\alpha}{2},1 + \pi K_f ,- \frac{{w^2 }}{{2z{\kern 1pt} }}} \right]
\label{4}
\end{equation}
\nobreak
where the function $_1 F_1$ is the confluent hypergeometric function of the first kind, and $\beta$  is a constant.  To satisfy the boundary condition
$\Gamma (r,0) = \Gamma _0 ' \,r^{ - 2 \pi K_i} $  at $t = 0$,
the large-$w$ expansion of the hypergeometric function requires $\alpha = 2 \pi K_i$\,, and
\begin{equation}
\beta  = \Gamma _0 ' \left( {\frac{{\:G(1 + \pi K_f  - \pi K_i )}}{{(2z)^{\pi K_i } \;G(1 + \pi K_f )}}} \right)
\label{5}
\end{equation}
where $G$ is the Euler gamma function.  The full solution for the distribution function then becomes
\begin{equation}
\Gamma (r,t) = \beta \,\,_1 F_1 \left[ {\pi K_i ,1 + \pi K_f , - \frac{{r^2 }}{{2z{\kern 1pt} \,t^{2/z} }}} \right]\;\,t^{ - 2\pi K_i /z} 
\label{6}
\end{equation}
which is Eq.\,(4) of the main text.
\subsection*{Scaling for quenches from $T_{KT}$}
\begin{figure}[ht]
\begin{center}
\includegraphics[width=0.5\textwidth]{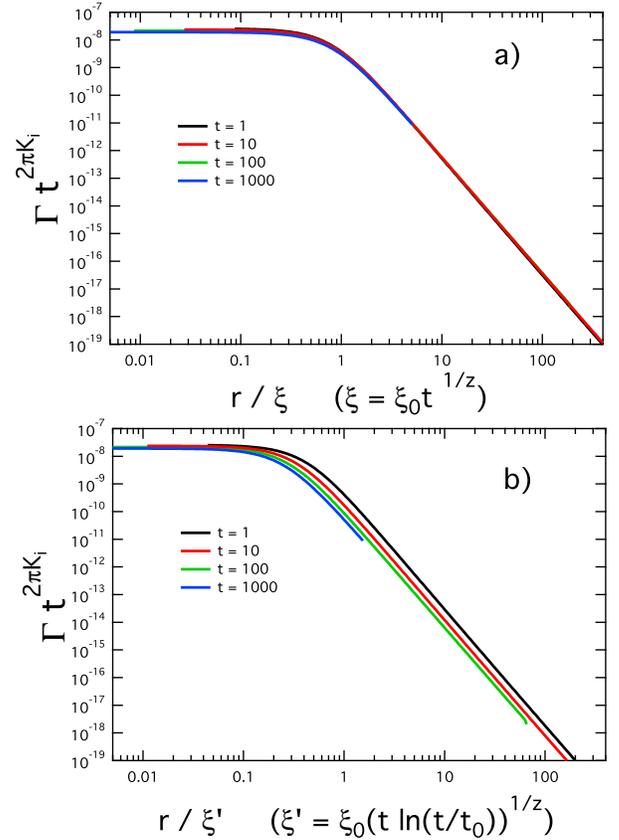}
\end{center}
\caption{Scaling plots of the vortex pair distribution for a quench from $T_{KT}$.   (Color online)}
\label{fig4}\end{figure}

Figure \ref{fig4}a shows the scaling plot of $t^{2\pi K_i /z} \,\Gamma$ versus $r / \xi$ for an instantaneous quench from $T_{KT}$ to 0.1 $T_{KT}$, using the computed values of $\Gamma$ from Eq.\,(2) of the main text. Here $\xi$ is taken to have the form $\xi = \xi_0 t^{1/z}$ with $\xi_0$ = 11.5, and $K_i$ is the value computed from Eq.\,(3) of the main text at the scale $r = \xi$.  It is seen that the curves collapse nearly completely, with only very slight deviations for $r / \xi \leq 1$, within the uncertainties of the value of $\xi_0$.  Figure \ref{fig4}b shows the same plot now using a log form for the dynamic scale, $\xi = \xi_0(t\ln(t/t_0))^{1/z}$, and taking $t_0$ = 0.02.  The curves clearly do not satisfy scaling, and the same holds true for all values of $t_0 < 1$.
\\*[6 cm]
\end{document}